\pdfoutput=1
\documentclass[aps,prb,showpacs,superscriptaddress,floatfix,twocolumn]{revtex4}
\usepackage{graphicx,epsf}
\usepackage{bm}
\usepackage{amssymb}
\usepackage{amsmath}

\def\Vec#1{{\bf #1}}

\begin{document}

\title{Correlated random hopping disorder in graphene at high magnetic fields:\\
Landau level broadening and localization properties
}

\author{A. L. C. Pereira} \affiliation{Faculdade de Ci\^encias
  Aplicadas, Universidade Estadual de Campinas, 13484-350 Limeira SP,
  Brazil}

\author{C. H. Lewenkopf} \affiliation{Instituto de F\'{\i}sica,
  Universidade Federal Fluminense, 24210-346 Niter\'oi RJ, Brazil}

\author{E. R. Mucciolo} \affiliation{Department of Physics, University
  of Central Florida, Orlando, FL 32816-2385, USA}

\date{\today}

\begin{abstract}
We study the density of states and localization properties of the
lowest Landau levels of graphene at high magnetic fields. We focus on
the effects caused by correlated long-range hopping disorder, which,
in exfoliated graphene, is induced by static ripples. We find that the
broadening of the lowest Landau level shrinks exponentially with
increasing disorder correlation length. At the same time, the
broadening grows linearly with magnetic field and with disorder
amplitudes. The lowest Landau level peak shows a robust splitting,
whose origin we identify as the breaking of the sublattice (valley)
degeneracy.
\end{abstract}

\maketitle

\section{Introduction}

The observation of the anomalous quantum Hall effect
\cite{novoselov05,zhang05} is one of the most striking and robust
manifestations of the underlying massless Dirac fermions in graphene
near half filling. The energy scales in graphene are such that the
quantization of the Hall plateaus can be observed even at room
temperature at sufficiently high magnetic fields.\cite{novoselov07}
The energetics in graphene also favors a direct experimental access to
the low-lying Landau levels by infrared spectroscopy
\cite{sadowski06,deacon07,jiang07a} and by scanning tunneling
spectroscopy,\cite{miller09,li-andrei09} something hardly possible in
conventional semiconductors.

Further information about the nature of the lowest Landau levels in
graphene has been recently obtained in thermal activation
experiments,\cite{jiang07b,giesbers07,giesbers09,giesbers09b,zeitler10}
showing that the zeroth Landau level $(n=0)$ is much sharper than the
first and higher Landau levels.
These observations are the main motivation of the theoretical study
presented in this paper.

Disorder is key to understand the electronic transport properties in
graphene,\cite{castroneto09,mucciolo10,abergel10,dassarma11}
particularly in the quantum Hall regime, where the conductivity
plateaus are conventionally explained by delocalized states surrounded
by localized ones. However, the mechanisms that lead to localization
in QHE in graphene are still not clear.\cite{martin09} Currently,
there is still some debate on the most relevant disorder mechanisms
for transport in graphene.\cite{mucciolo10} Among those, ripple
disorder is believed to play an important role. Static ripples give
rise to random correlated hopping disorder,\cite{guinea08,vozmediano10}
which is the disorder mechanism analyzed in this paper.


The shape and width of the lowest Landau levels (LLs) in graphene have
been investigated in several theoretical
studies.\cite{koshino07,pereira08,zhu09,pereira09,kawarabayashi09a,kawarabayashi10,zhu10,peeters10,peeters10b}
The broadening of the LLs differs among disorder models. In
particular, the inclusion of a finite correlation length on the
hopping (off-diagonal) disorder model was recently reported to induce
an anomalously sharp $n=0$ LL compared to higher
levels.\cite{kawarabayashi09a,kawarabayashi10} It was found that the
width of the zeroth LL ($\Gamma_0$) shrinks to zero as soon as the
hopping correlation length $\lambda$ exceeds the lattice parameter
$a$, in line with analytical studies of the effects of long-range
chiral disorder.\cite{ostrovsky08}

Regarding the localization properties of the lowest LL, numerical
simulations using uncorrelated hopping disorder
\cite{koshino07,pereira09} and white-noise random magnetic flux
disorder \cite{schweitzer08,schweitzer09} observe an interesting
distinct qualitative feature in the quantum Hall spectrum of graphene,
namely, a splitting. It was found that, in such chiral disorder
models, the lowest LL splits into two Gaussian shaped peaks, even in
the absence of both a Zeeman term and electron-electron interactions.
The splitting energy $\Delta E$ is linearly proportional to the
disorder strength and scales with the square root of the applied
perpendicular magnetic field.\cite{schweitzer08,pereira09} A similar
square root magnetic field dependence of the splitting of the $n=0$
Landau level has been experimentally observed in
Ref.~\onlinecite{jiang07b}.

In this paper, instead of the white-noise random hopping (or magnetic
flux) model, we address the more realistic correlated random hopping
disorder model.\cite{kawarabayashi09a,kawarabayashi10} We present a
systematic study of the shape of the lowest LLs and their localization
properties as a function of the hopping disorder correlation length,
as well as of other relevant parameters of the system, such as
disorder amplitude and magnetic field. We find that $\Gamma_0$ decays
exponentially with the correlation length, never fully vanishing for
any finite $\lambda$. More importantly, we observe that the ratio
$\Gamma_1/\Gamma_0$ depends only on the disorder correlation length,
showing no significant variation neither with the disorder strength,
nor with the magnitude of the applied magnetic field $B$, provided the
system is in the quantum Hall regime. In addition, we study the
splitting of the $n=0$ LL, which is inferred from the analysis of the
participation ratio. We show that this splitting shrinks with
increasing values of $\lambda$, but is still present even for
correlation lengths for which the $n=0$ LL width becomes very small.

The paper is organized as follows. In Sec. \ref{sec:model}, we present
the model used in our numerical simulations. The analytical framework
for the interpretation of our results is discussed in
Sec.~\ref{sec:dirac}. Next, we analyze the spectral,
Sec.~\ref{sec:spectrum}, and localization properties,
Sec.~\ref{sec:localization}, of the model. We conclude summarizing our
results and discussing their relevance to the interpretation of
experiments on the quantum Hall effect in graphene in
Sec.~\ref{sec:conclusions}.

\section{Model description}
\label{sec:model}

Graphene is a monolayer honeycomb lattice of carbon atoms with a
lattice constant $a=2.46$~\AA. Its primitive unit cell contains a pair
of atoms that form two triangular sublattices, denoted by $A$ and
$B$. The tight-binding Hamiltonian model for a graphene monolayer
reads
\begin{equation}
\label{eq:tb_hamiltonian}
H = \sum_{\left<ij\right>} \left( t_{ij}e^{i\phi_{ij}}\, c_{i}^{\dagger}
c_{j}{} + \rm{H.c.} \right),
\end{equation}
where the sum runs over nearest-neighbor sites. The external magnetic
field $B$, perpendicular to the graphene sheet, is included by
Peierls' substitution, namely, $\phi_{ij}= 2\pi(e/h) \int_{{\bm
    r}_j}^{{\bm r}_i} \bm{A} \! \cdot \!  d \bm{l} \;$. In the Landau
gauge, $\bm{A}=(0, Bx)$ and considering a brick wall lattice, which is
topologically equivalent to the hexagonal lattice,\cite{wakabayashi99}
one has $\phi_{ij}\!=\!0$ along the $x$ direction and $\phi_{ij}\!=\pm
\pi (x/a) \Phi / \Phi_{0}$ along the $\mp y$ direction, with $\Phi /
\Phi_{0}=Ba^{2}\sqrt{3}e/(2h)$ per unit cell.

\begin{figure}[b]
\includegraphics[width=0.96\columnwidth]{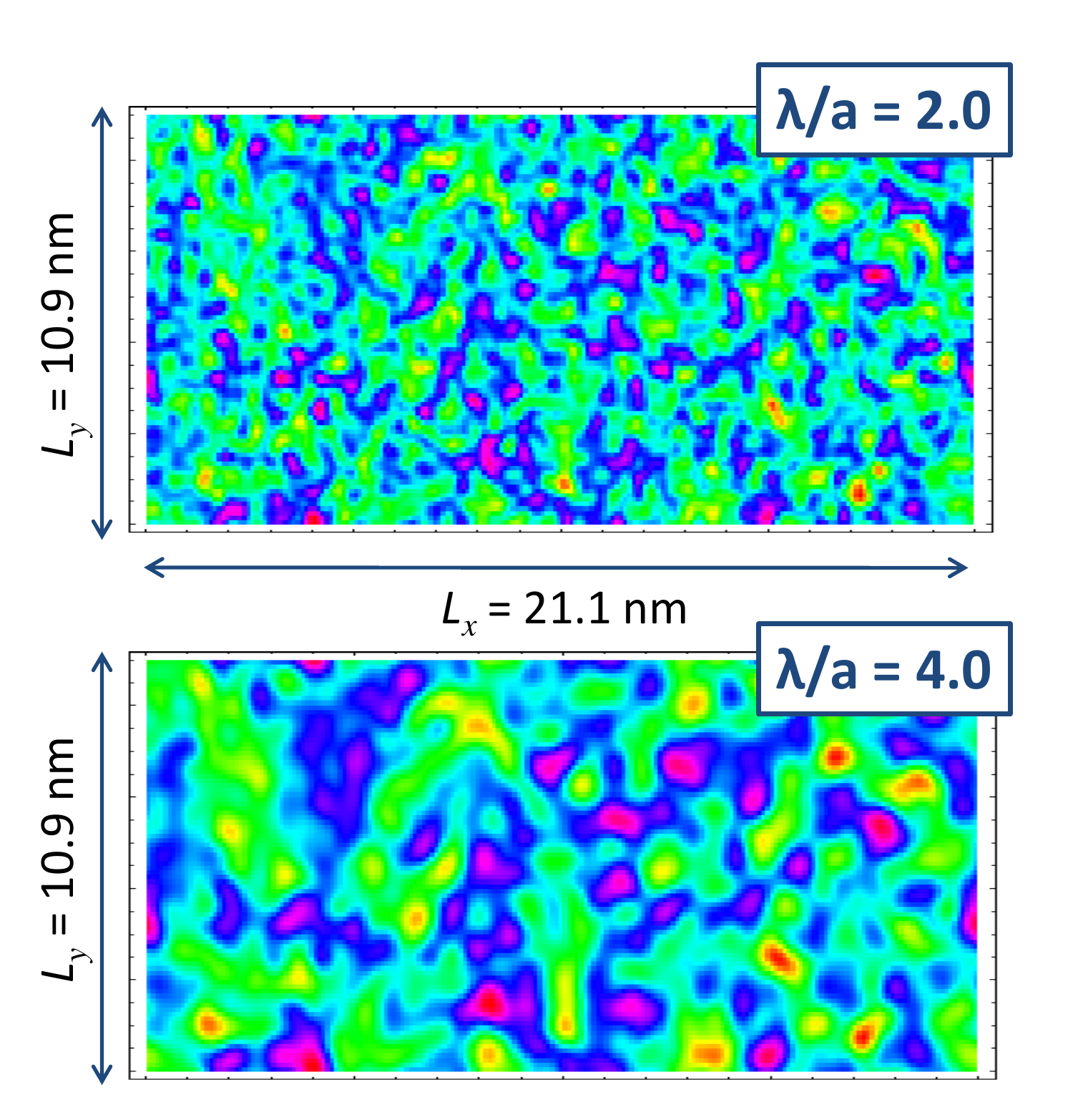}
\caption{Typical spatial ``landscapes'' of the fluctuation in the
  hopping matrix elements $\delta t_{ij}/W=(t_{ij} - t)/W$ for two
  different correlation lengths ($\lambda=2a, 4a$). The color scale
  represents the magnitude of $\delta t_{ij}/W$. The lattices shown
  above have $M$$\times$$N$ = 100$\times$90 atoms, which corresponds
  to lateral dimensions $L_x$$\times$$L_y$ = 21.1 nm $\times$ 10.9
  nm.}
\label{fig:t_landscape}
\end{figure}

The random hopping disorder is implemented by randomly choosing the
hopping parameters $t_{ij}$ from a uniform distribution of width $W$
around the average value $t=2.7$ eV. In addition, we impose here a
spatial correlation to the hoppings following a Gaussian profile of
width $\lambda$. Figure \ref{fig:t_landscape} illustrates typical
realizations of the disordered hopping parameter $t_{ij}$ for two
different correlation lengths ($\lambda=2a,4a$). The color scale
refers to the hopping amplitude at the middle point between the two
nearest-neighbor sites $i$ and $j$. As expected,
Fig. \ref{fig:t_landscape} shows smoother (less abrupt) variations of
the hopping values with increasing correlation length.

We consider graphene lattices of $M$$\times$$N$ carbon atoms ($M$
zigzag chains, each containing $N$ atoms) with periodic boundary
conditions. The linear sizes of the lattices are
\begin{equation}
 L_x=(M-1)\frac{\sqrt{3}}{2}a, \;\;\;\;\,\,\,\,\, L_y=(N-1)\frac{a}{2}.
\end{equation}

Most of the numerical results shown in this paper are calculated for
lattices of $M$$\times$$N$ = 100$\times$90 atoms, corresponding to a
size $L_x$$\times$$L_y$ = 21.1 nm $\times$ 10.9 nm, the same lattice
dimensions shown in Fig. \ref{fig:t_landscape}.

\section{Dirac Hamiltonian: analytical results}
\label{sec:dirac}

Near half filling, the low-energy properties of the tight-binding
Hamiltonian in Eq.~\eqref{eq:tb_hamiltonian} are described by
noninteracting massless Dirac fermions in an uniform perpendicular
magnetic field, with an effective Hamiltonian given by
\cite{shon98,zheng02}
\begin{equation}
 H_0 = v_F
\begin{pmatrix}
 0 & \pi_x - i \pi_y & 0 & 0 \\
 \pi_x + i \pi_y & 0 & 0 & 0 \\
 0 & 0 & 0 & \pi_x + i \pi_y \\
 0 & 0 & \pi_x - i \pi_y & 0
\end{pmatrix},
\label{eq:H_Dirac}
\end{equation}
where $\hbar v_F = \sqrt{3}a t /2$ is the Fermi velocity, $\bm{\pi}=
\bm{p} + e\bm{A}/c$, and $\bm{p}$ stands for the electron momentum
operator. The Hamiltonian \eqref{eq:H_Dirac} operates on a
four-components wave function
$(\Psi^K_A,\Psi^K_B,\Psi^{K'}_A,\Psi^{K'}_B)$, where $\Psi^K_A$ and
$\Psi^K_B$ represent the envelope functions at $A$ and $B$ sites for
the $K$ point and $\Psi^{K'}_A$ and $\Psi^{K'}_B$ for $K'$,
respectively. In this paper we do not consider explicitly the electron
spin degree of freedom and all states are assumed spin degenerate.

The eigenstates of the Hamiltonian \eqref{eq:H_Dirac} are labeled by
$(j,n,k)$, with the valley index $j=K,K'$, the Landau level index $n =
0,\pm 1, \dots$, and the wave vector $k$ along $y$
direction.\cite{shon98} The eigenenergy depends solely on $n$ as
$\varepsilon_n = \hbar\omega_B\,\,{\rm sgn}(n)\sqrt{|n|}$, where
$\hbar\omega_B = \sqrt{2}\gamma/\ell_B$ with the magnetic length given
by $\ell_B = \sqrt{\hbar/eB}$. For $\Phi / \Phi_{0} < 0.05$, which
implies $\ell_B/a \gg 1$, lattice size effects have a negligible
influence on the graphene LLs in this model.

The eigenfunctions are written as
\begin{eqnarray}
 \Psi^K_{nk} &=& \frac{C_n}{\sqrt{L}}\exp(iky)
\left(
\begin{array}{c}
{\rm sgn}(n) (-i) \phi_{|n|-1,k}\\
\phi_{|n|,k}\\0\\0
\end{array}
\right), \\
 \Psi^{K'}_{nk} &=& \frac{C_n}{\sqrt{L}}\exp(iky)
\left(
\begin{array}{c}
0\\0\\
\phi_{|n|,k}\\
{\rm sgn}(n) (-i) \phi_{|n|-1,k}
\end{array}
\right).
\label{eq_LL}
\end{eqnarray}
Here $C_n=1$ for $n=0$, $C_n=1/\sqrt{2}$ for $n\neq 0$, and
\begin{equation}
\phi_{n,k}(x) = (2^{n}n!\sqrt{\pi}\ell_B)^{-1/2} \,\,e^{-z^2/2}H_{n}(z),
\end{equation}
with $z = (x-k\ell_B^2)/\ell_B$ and $H_n(z)$ denoting Hermite
polynomials.

It was realized early \cite{koshino07} that the level with $n=0$ is
special since its amplitude is non-zero only in one of the
sublattices, namely, at $B$ sites for $K$ and $A$ sites for
$K^\prime$. Consequently, while a random on-site disorder potential
gives only intravalley mixing within either the $K$ and $K^\prime$
valleys, random hopping causes intervalley mixing. (Notice that this
is quite the opposite of what occurs at zero magnetic field when
diagonal disorder is present.) The wave function in LLs with $n\neq 0$
has nonzero amplitudes on both $A$ and $B$ sites, so that intervalley
mixing is always possible.

In this study, we consider hopping disorder caused by randomness in
the hopping integral connecting neighboring $A$ and $B$ sites. This
disorder can be long ranged, when caused by static ripples, or short
ranged, when originated by scatterers located at points in-between
neighboring sites.

Assuming that $t$ shifts to $t + \delta t$ between neighboring sites
${\bm R}_A$ and ${\bm R}_B$, hopping disorder gives rise to a
short-range potential given by\cite{koshino07}
\begin{equation}
 U({\bm r}) =
\left(
\begin{array}{cccc}
 0 & z_A^*z_B & 0 & z_A^*z_B^\prime \\
 z_B^*z_A & 0 & z_B^*z_A^\prime & 0 \\
 0 & z_A^{\prime*} z_B & 0 & z_A^{\prime*} z_B^{\prime} \\
 z_B^{\prime*} z_A & 0 & {z_B^\prime}^* z_A^{\prime} & 0
\end{array}
\right)
u_{\rm h} \delta(\Vec{r}-\Vec{R}_i),
\end{equation}
with $u_{\rm h} = (\sqrt{3}a^2/2)\delta t$, $z_X =
e^{i\Vec{K}\cdot\Vec{R}_{X}}$, and $z'_X =
e^{i\Vec{K}'\cdot\Vec{R}_{X}}$ for $X = A$ and $B$. For a hopping
disorder concentration $n_{\rm h}$, the self-consistent Born
approximation estimates the Landau level broadening as $\Gamma_{\rm
  short} = (2 n_{\rm h})^{1/2} u_{\rm h}/(\pi \ell_B)$, independent of
Landau level index. Numerical simulations \cite{pereira09} exhibit the
same scaling of $\Gamma_{\rm short}$ with $\sqrt{B}$, but show that
$\Gamma_{\rm short}$ increases with the index $n$.

Alternatively, when the randomness in the hopping integrals shows
long-range correlations, the disorder Hamiltonian can be formulated in
terms of an effective random magnetic
field.\cite{abanin07_XY,vozmediano10} In this case, one assumes that
lattice deformations cause a smooth shift in the hopping integrals
between the site $j$ and its three nearest neighbors $i$. At low
energies, this effect can be incorporated into Dirac Hamiltonian by
introducing an effective vector potential that reads
\begin{equation}
A^{\rm eff}_\pm ({\bm r}_j) = \frac{c}{e} \sum_{i=1}^3 \delta t_i
({\bm r}_j)e^{\pm i {\bm q}_0\cdot \hat{{\bm e}}_i}.
\end{equation}
Here, $A^{\rm eff}_\pm = A^{\rm eff}_x \mp i A^{\rm eff}_y$, ${\bm
  e}_i$ are vector connecting a lattice site to its neighbors, and
${\bm q} \approx \pm {\bm q}_0$, where ${\bm q}_0 = 4\pi/(3\sqrt{3}a)
{\bm e}_y$ with the Cartesian coordinate $x$ running along the
armchair direction. The subscript $+\ (-)$ corresponds to the
$K\ (K')$ valley. Random hopping is accounted for by adding the
effective vector potential ${\bm A}^{\rm eff}$ to the momentum
operator $\bm{\pi}$ appearing in Eq.~\eqref{eq:H_Dirac},
namely,\cite{abanin07_XY}
\begin{equation}
 H_\pm = v_F
\begin{pmatrix}
 0 & \pi_x - i \pi_y + \frac{e}{c}A_\pm^{\rm eff}\\
 \pi_x + i \pi_y + \frac{e}{c}A_\pm ^{\rm eff *}& 0
\end{pmatrix}.
\label{eq:H_Dirac-randomB}
\end{equation}
Notice that the effect of $A^{\rm eff}_\pm$ is to locally shift the
Dirac cones $K$ and $K^\prime$ in opposite directions and there is no
valley mixing. The structure of $H_\pm$ immediately reveals that the
$n=0$ states are unique: Since $\langle \Psi^K_{0k} | H - H_0 |
\Psi^{K^\prime}_{0k^\prime} \rangle =0$, in lowest order, long-range
hopping disorder does not affect the $n=0$ states.

Unfortunately, there is no such simple picture for treating the
crossover regime between short- and long-range hopping disorder. In
the following, we interpret the results of our numerical simulations
by invoking the pure long-range description provided by
Eq.~\eqref{eq:H_Dirac-randomB}, what is known about short-range
hopping disorder, and by building a plausible interpolation between
these two limits.

\section{Spectral properties}
\label{sec:spectrum}

In this Section, we analyze the spectral properties of graphene
obtained from the tight-binding model with correlated random hopping
presented in Sec.~\ref{sec:model}. We will focus our attention on the
width of the disorder-broadened Landau levels and its dependence on
the disorder correlation length $\lambda$, on the disorder amplitude
$W$, on the magnetic flux $\Phi$, and on the LL index $n$.

\begin{figure}[ht]
\includegraphics[width=0.96\columnwidth]{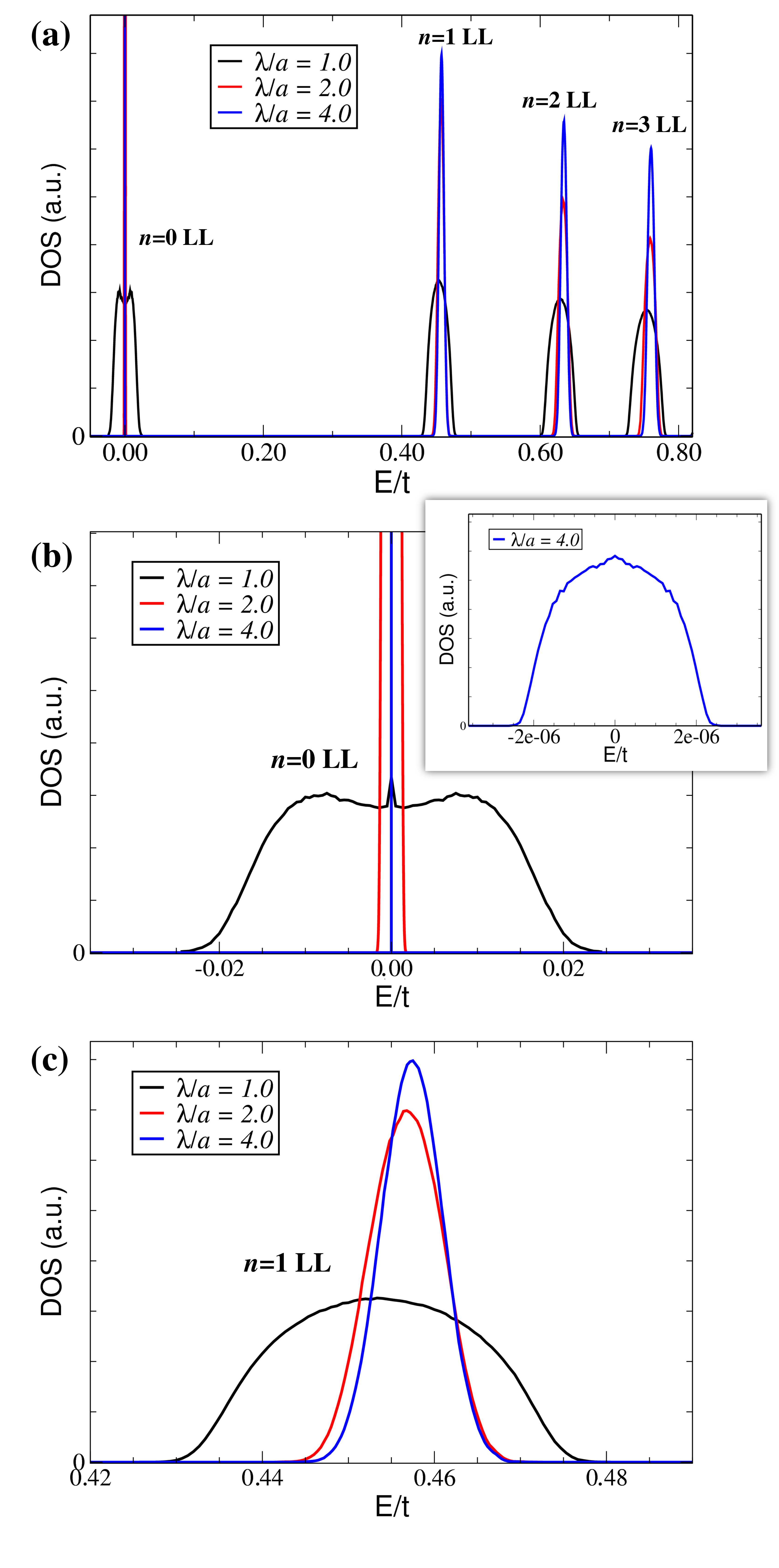}\\
\caption{Density of states (in arbitrary units) for a magnetic flux
  $\Phi /\Phi_0$=$0.02$ for three different correlation lengths
  $\lambda/a$ of the random hopping potential (in all cases $W/t=0.3$
  and the lattice size is $M$$\times$$N$=100$\times$90 sites). (a) DOS
  showing the $n$=0, 1, 2, and 3 Landau bands. (b) Zoom of the $n=0$
  LL states. (c) Zoom of the $n=1$ LL states.}
\label{fig:DOS}
\end{figure}

Figure \ref{fig:DOS}a shows the density of states (DOS) corresponding
to the four lowest LLs ($n=0, 1, 2$, and $3$) broadened by a
correlated random hopping disorder of amplitude $W/t=0.3$ for
different correlation lengths, namely, $\lambda/a$ = 1, 2, and
4. Since particle-hole symmetry is preserved by the nearest-neighbor
hopping disorder model, we only show the $n \ge 0$ LLs. The magnetic
flux considered is $\Phi / \Phi_{0}$=0.02.
The results typically correspond to averages over 600 disorder
realizations.

Figures \ref{fig:DOS}b and \ref{fig:DOS}c are zooms of the DOS around
the $n$=0 and the $n$=1 LLs, respectively, indicating that the
broadening shrinks for increasing values of $\lambda$. An inspection
of Fig.~\ref{fig:DOS}c shows that, for the $n=1$ LL, an increase in
the correlation length causes only small modifications on the shape of
this Landau band, plus an overall reduction of the width
$\Gamma_1$. Higher Landau levels, $n>1$, show the same qualitative
behavior as $n=1$. In contrast, for the $n$=0 LL
(Fig. \ref{fig:DOS}b), one observes a much stronger suppression of the
level broadening upon increasing $\lambda$, in agreement with
numerical results obtained in
Ref.~\onlinecite{kawarabayashi09a}. However, the inset of Fig.
\ref{fig:DOS}b shows that, despite the large reduction of $\Gamma_0$
with increasing $\lambda$, the width never goes to zero within the
range of $\lambda$ we used. This is at odds with the results reported
in Ref.~\onlinecite{kawarabayashi09a}, where an abrupt transition to
zero width was observed. It is worth mentioning that, when $\lambda$
increases, making the width of the $n=0$ LL much smaller than the
width of the higher levels, it is important to use DOS histograms with
a much finer energy resolution around the $n=0$ LL than for the higher
LLs. Not doing so can lead to an erroneous impression that the $n=$ LL
width vanishes sharply as the correlation length increases.

In the following, we describe how we define and quantify the width
$\Gamma_n$ of the disordered-broadened LLs. For the $n>0$ LLs, which
display a Gaussian-like shape, $\Gamma_n$ is taken as the full width
at half-height. Figure \ref{fig:DOS} clearly shows that the DOS shape
of the $n=0$ LL is very different from the higher LLs. As pointed out
in Ref.~\onlinecite{pereira09}, an off-diagonal disorder model induces
a splitting of the $n=0$ LL into two degeneracy-broken $n=0$ Landau
bands, causing the observed DOS shape for the $n=0$ LL, namely, not
fully split levels. This is so because the energy splitting always has
the same order of magnitude of the LL broadening. Here, the $n=0$ LL
shapes observed in Fig. \ref{fig:DOS}b can be reasonably well fitted
by a superposition of two equal Gaussian curves. Therefore, the $n=0$
LL width is considered as the width at half-height of one of these
superposed bands (which is approximately half the width at half-height
of the entire band).

\begin{figure}[t]
\includegraphics[width=\columnwidth]{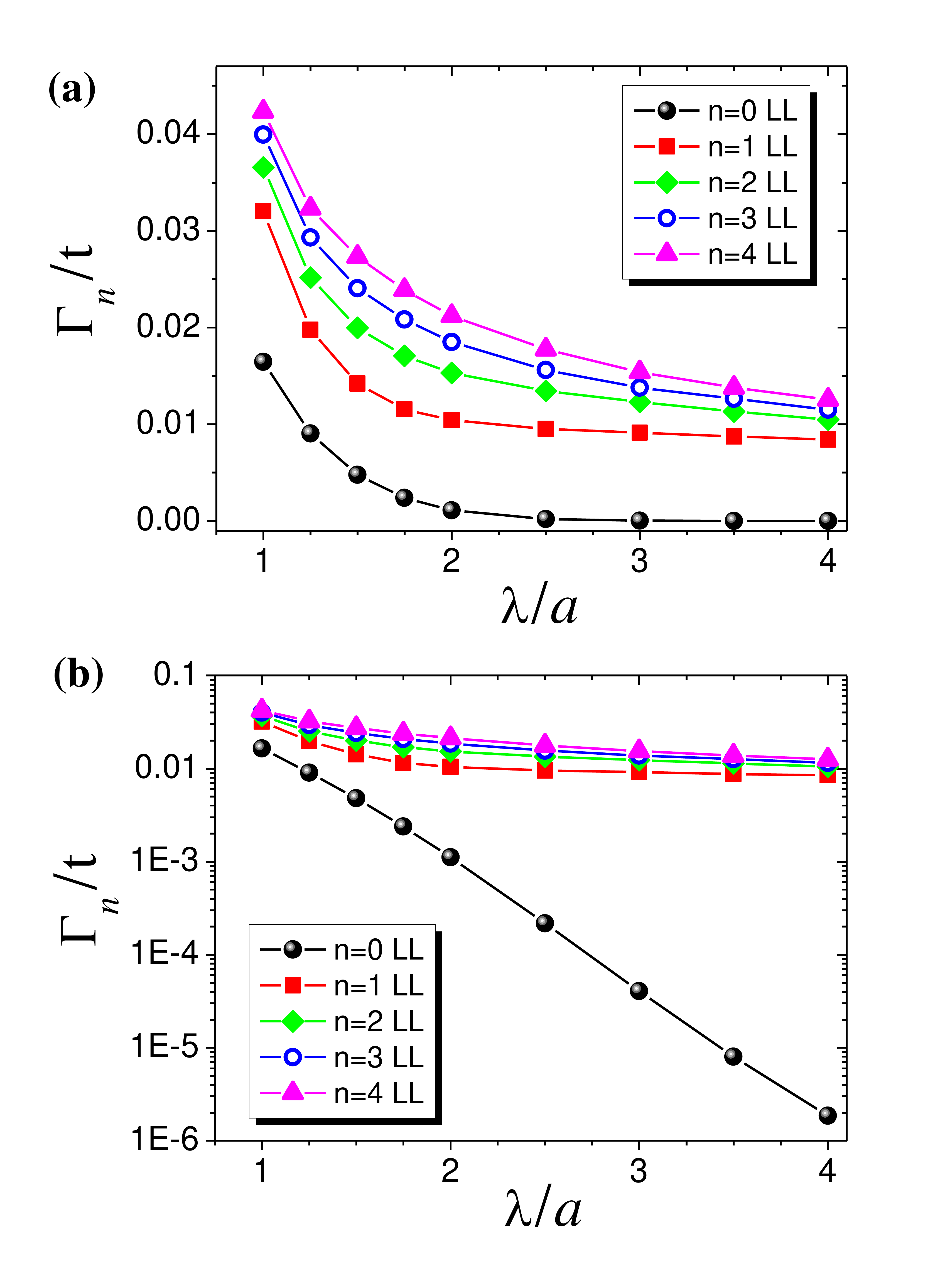}
\vskip-0.6cm
\caption{Width of the $n$th Landau level $\Gamma_n/t$, $n=0,1,2,3,4$,
  as a function of the disorder correlation length $\lambda/a$ for
  $\Phi/\Phi_0=0.02$ and $W/t=0.3$. (a) Linear scale. (b) Log-linear
  scale.}
\label{fig:LLwidth}
\end{figure}

Figure \ref{fig:LLwidth} shows how the LL widths decrease as the
hopping disorder correlation length grows. In Fig.~\ref{fig:LLwidth}a,
one observes that not only $\Gamma_0$ decreases with $\lambda$, but
all other LLs do so. When examining the same data in a log-linear
graph (Fig. \ref{fig:LLwidth}b), one observes that the $n=0$ LL
behaves quite differently from the others. For $n>0$, the widths
$\Gamma_n$ decrease slowly with increasing $\lambda/a$. Figure
\ref{fig:LLwidth} shows an apparent tendency to saturation at a value
$\Gamma_n/t \approx 0.01$. We computed $\Gamma_{n>0}$ for larger
values of $\lambda/a$ (not shown here) and concluded that this in not
quite correct. Instead, we observe that the rate by which all
$\Gamma_{n>0}$ decrease becomes smaller as $\lambda/a$ becomes
larger. The behavior is very different for the $n$=0 LL, whose width
decays exponentially with $\lambda$. This decay is sustained down to
the numerical precision of our simulations.

Further insight about the effect of long-range correlated hopping
disorder on the levels widths $\Gamma_n$ can be gained from a
perturbation theory analysis. Let us denote the disorder gauge
potential by $V=H_\pm-H_0$, where both $H$ and $H_0$ are defined in
Sec.~\ref{sec:dirac}.

In first order, the matrix elements required to calculate the energy
corrections for the degenerate states that belong to the $n$th Landau
level at the $K$ valley are
\begin{equation}
V_{nK,kk^\prime}^{(1)} = \langle \Psi_{n'k^\prime}^K|V|\Psi_{nk}^K\rangle.
\end{equation}
Since long-range hopping disorder does not mix valleys, $K$ is a good
quantum number in this model.  As discussed in Sec.~\ref{sec:model},
$E_{0}^{(1)}=0$. For $n>0$, the situation is different. Exact
diagonalization in a $(n>0, K)$ subspace involves matrix elements of
the kind $V_{n\neq 0,K;kk^\prime}^{(1)}$. Due to the spinor structure
of $\Psi_{nk}^K$, the evaluation of such matrix elements amounts to
the spatial integration of the product $\exp[i(k-k^\prime)y]
\phi_{nk'}(x)\phi_{n-1, k}(x)$ times the Gaussian fluctuating gauge
potential. This results in non-zero matrix elements, but they are very
quickly suppressed as $\lambda/\ell_B$ becomes large.

Let us define
\begin{equation}
V_{nK,kk^\prime}^{(2)} = {\sum_{n^\prime\neq n}} \frac {|\langle
  \Psi_{n^\prime k^\prime}^K|V|\Psi_{nk}^K\rangle|^2}{\hbar
  \omega_B(\sqrt{n^\prime}-\sqrt{n^{}})}
\end{equation}
to help us to discuss second-order effects. The evaluation of
$V_{n=0,K;kk^\prime}^{(2)} $ involves matrix elements where products
of $\phi_{1,k^\prime}$ and $\phi_{0,k}$ appear. For long-range
disorder, such matrix elements vanish with increasing
$\lambda/\ell_B$. In summary, the perturbation fails to mix states
within the $n=0$ multiplet and also with $n^\prime \neq 0$. Hence, we
expect the same behavior at all order of perturbation theory. This is
consistent with statement that $\Gamma_0=0$ for long-range hopping
disorder.\cite{ostrovsky08} Using the reasoning presented above,
matrix elements of the kind $V_{n\neq0,K;kk^\prime}^{(2)}$ involve,
among other components, the integration of products wave function
amplitudes such as $\exp[i(k-k^\prime)y]\phi_{nk'}(x)\phi_{n,
  k}(x)$. These become small for $k\neq k^\prime$ in the limit of
$\lambda/\ell_B \gg 1$ and, for $k=k^\prime$, are quite independent of
$\lambda/\ell_B \gg 1$.

This analysis rules out long-range hopping disorder as the mechanism
behind the exponential suppression of $\Gamma_0$ with increasing
$\lambda$. We speculate that this behavior is caused by matrix
elements that admix valleys, a remnant of the crossover regime. As a
consequence, we expect $\Gamma_0$ to scale linearly with the disorder
strength $W$ and the eigenstates to be a superposition of a
wave functions with probability amplitudes in both sublattices. Our
simulations are in line with the latter statement, as we discuss
below.

\begin{figure}[b]
\vskip-0.2cm
\includegraphics[width=\columnwidth]{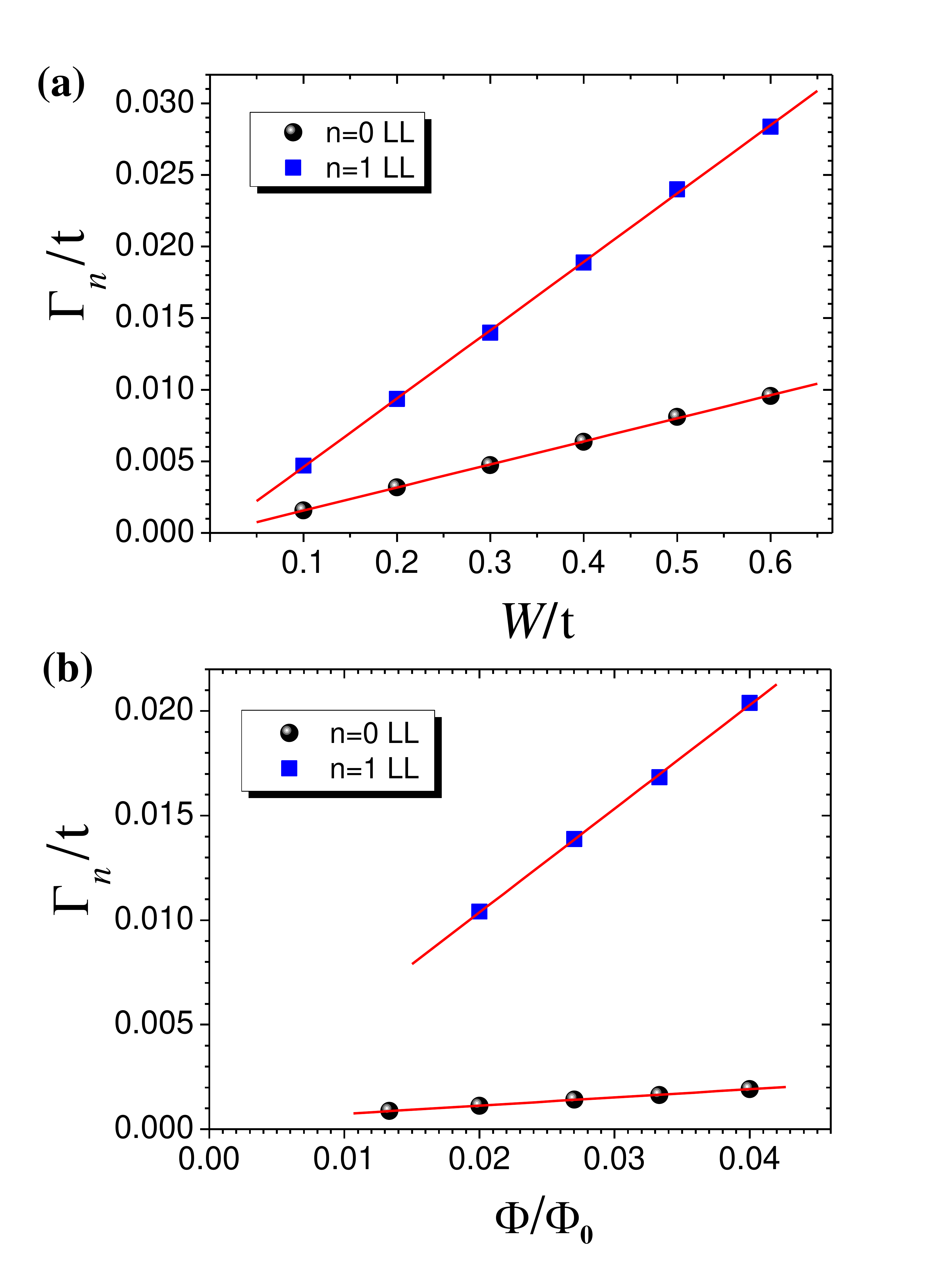}
\vskip-0.5cm
\caption{(a) Landau level width $\Gamma_n/t$ as a function of the
  hopping disorder strength $W/t$ for the n=0 and n=1 LLs (magnetic
  flux fixed at $\Phi/\Phi_0$=$0.02$ and $\lambda/a$=$1.5$). (b)
  $\Gamma_n$ as a function of magnetic flux $\Phi/\Phi_0$ for the same
  LLs (now for a fixed disorder strength $W/t$=$0.3$ and
  $\lambda/a$=$2.0$).}
\label{fig:Gamma_vs_flux}
\end{figure}

When a magnetic field is present, there is an important length scale
to be considered, namely, the magnetic length; for convenience, let us
write it in the form $\ell_B/a$=0.371$/\sqrt{\Phi/\Phi_0}$. For the
magnetic flux used in our simulations ($\Phi/\Phi_0$=$0.02$), we
obtain $\ell_B/a$=$2.62$, which is close to the values of $\lambda/a$
used as well.
Therefore, we expect the magnetic length to play a role in any
interpretation of our results. Indeed, Fig.~\ref{fig:LLwidth}a
suggests that the LL widths change its dependence with $\lambda$ for
$\lambda \agt \ell_B$. However, Fig.~\ref{fig:LLwidth}b makes clear
that a slow dependence on $\lambda$ only occurs for the $n>$0 LL. This
is in line with our perturbative analysis. Unfortunately, the picture
is not entirely consistent: We expect the second-order terms to be
dominant in the calculation of the broadening of $\Gamma_{n>0}$ for
$\lambda/\ell_B$, which is not observed in the simulations (see
discussion below). This remains to be understood.

We call attention to the fact that the authors of
Ref.~\onlinecite{kawarabayashi09a} used the same flux value we
considered here, as well as a similar range for the correlation length
$\lambda$.
However, in our calculations, a finite width of the $n=0$ LL can be
seen even for $\lambda/a$=$4$ within the numerical precision we use.
We checked (not shown here) that these results are not influenced by
varying the system sizes, i.e, there are no finite lattice-size
effects in the parameter region we investigated.

We have also investigated the dependence of the $n=0$ and $n=1$ LLs
widths on the disorder and magnetic field amplitudes (Fig.
\ref{fig:Gamma_vs_flux}). For both parameters, there is a clear linear
increase.

\begin{figure}[t]
\includegraphics[width=1.03\columnwidth]{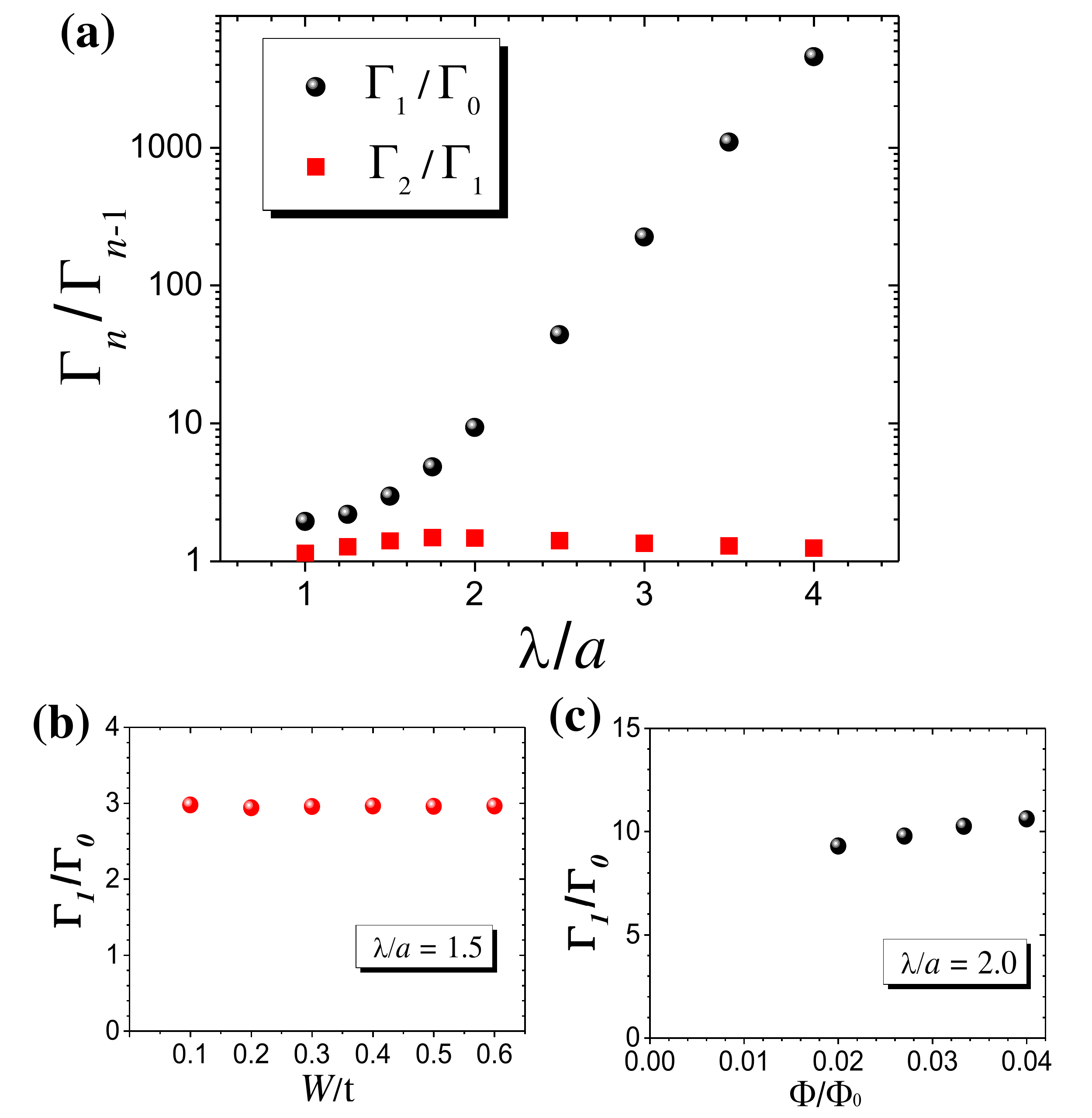}\\
\caption{(a) Ratio between different Landau level widths
  $\Gamma_n/\Gamma_{n-1}$ as a function of $\lambda/a$. The ratio
  between higher LLs widths, like $\Gamma_2/\Gamma_1$, is quite
  independent of the correlation length $\lambda$. In contrast, the ratio
  $\Gamma_1/\Gamma_0$ grows rapidly with $\lambda/a$ (notice the
  logarithmic scale). (b) Ratio $\Gamma_1/\Gamma_0$ as a function of the disorder strength $W/t$ for
  $\lambda/a=1.5$. (c) Ratio $\Gamma_1/\Gamma_0$ as a function of the magnetic flux  $\Phi/\Phi_0$
  for $\lambda/a=2.0$. The panels (b) and (c) show that $\Gamma_1/\Gamma_0$
  is almost independent of $W/t$ and $\Phi/\Phi_0$. In the log-linear plot of panel (a) the results corresponding
  to different values of $W/t$ and $\Phi/\Phi_0$ collapse under the same dot. }
\label{fig:universal}
\end{figure}

Since the LL widths depend linearly on $W$ and on $\Phi/\Phi_0$, we
conclude that there is universality in the behavior of the Landau
level widths, namely, the ratio between different Landau level widths,
$\Gamma_n/\Gamma_{n^\prime}$, depends solely on $\lambda/a$. This is
illustrated in Fig.~\ref{fig:universal}a, where one can see the ratio
$\Gamma_1/\Gamma_0$ growing rapidly with $\lambda/a$ (notice the
logarithmic scale), while $\Gamma_2/\Gamma_1$ remains essentially
constant. This novel result allows our simulations to make contact
with the experiments. Notice that due to computational limitations,
our lattices sizes constrain us to consider unrealistically large
values of magnetic field magnitudes. However, since the ratios of LL
widths are rather insensitive to the values of $W$ and $\Phi/\Phi_0$,
as illustrated by Figs.~\ref{fig:universal}b and c,
we expect this result to apply for realistic settings as well.

\begin{figure}[t]
\vskip-0.2cm
\includegraphics[width=.98\columnwidth]{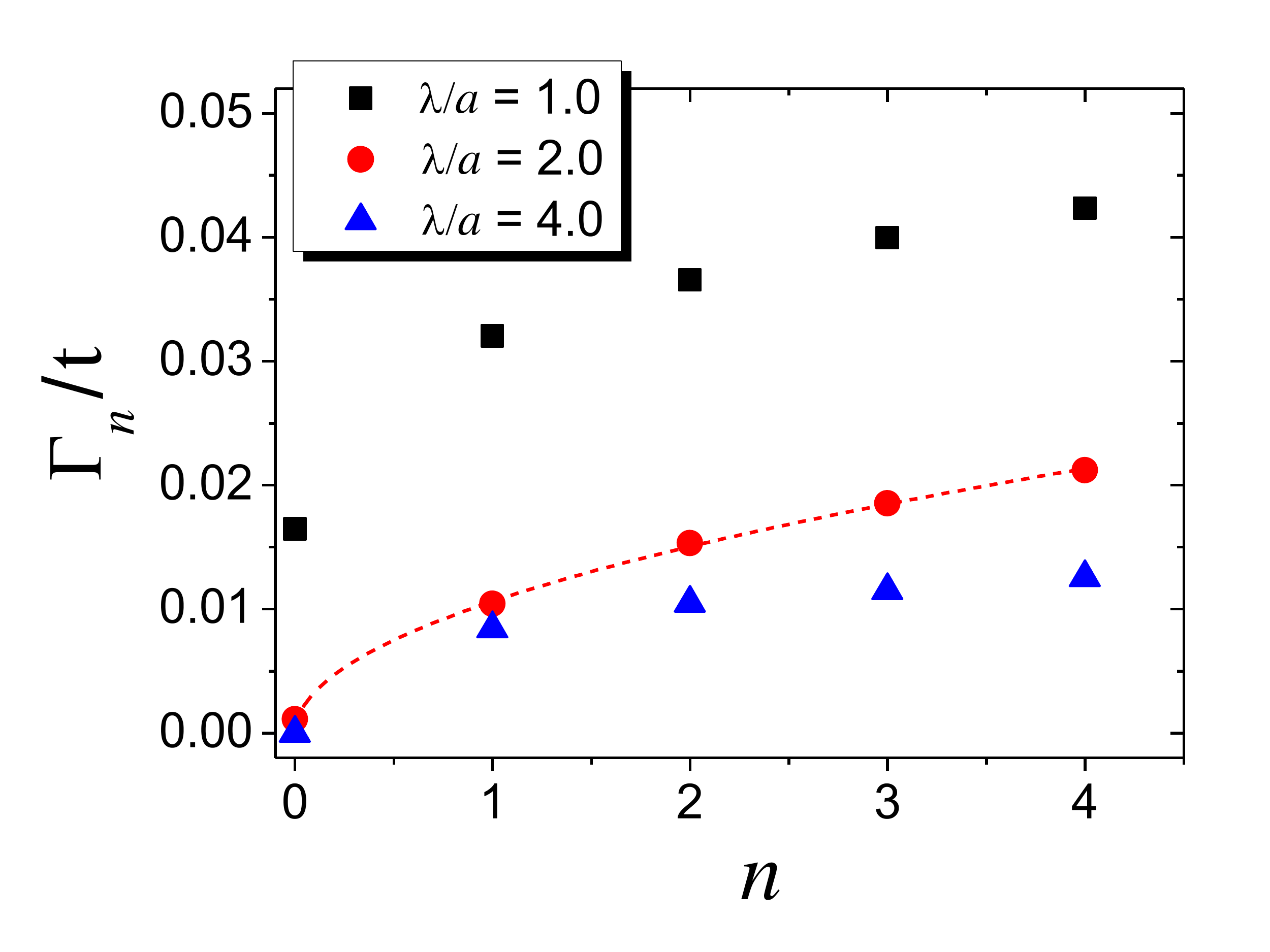}
\vskip-0.5cm
\caption{Landau level width $\Gamma_n/t$ as a function of LL index $n$
  for three different values of $\lambda/a$, considering
  $\Phi/\Phi_0=0.02$ and $W/t=0.3$ (same magnetic flux, disorder
  amplitude, and lattice sizes considered in
  Fig. \ref{fig:LLwidth}). The dashed line corresponds to a fitting
  with square root dependence, which fits well only the $\lambda/a =
  2$ case (see text).}
\label{fig:Gamma_vs_n}
\end{figure}

The dependence of the width on the LL index is shown in
Fig. \ref{fig:Gamma_vs_n}. In this case as well, $\Gamma_n$ increases
with $n$ when $\lambda$, $W$, and $\Phi$ are fixed (notice that the
values for these parameters are the same considered in
Fig. \ref{fig:LLwidth}). We find that for $\lambda/a=2$, the
$\Gamma_n$ versus $n$ curve is perfectly fitted by the functional form
$y=A\sqrt{x}$, with $A$=0.0107 (dashed line in
Fig. \ref{fig:Gamma_vs_n}). For other values of $\lambda$, including
the cases $\lambda/a=1$ and $\lambda/a=4$, the numerical data do not
show a square root dependence on LL index $n$. At this point, it is
worth comparing these results to the effect of a diagonal disorder on
the DOS in graphene at the quantum Hall regime. The Landau level
broadening $\Gamma_n$ for a diagonal white-noise disorder is quite
independent of the LL index $n$, \cite{pereira08,zhu09} while for the
correlated diagonal disorder, $\Gamma_n$ is observed to slightly
decrease with increasing $n$.\cite{pereira08} These effects for the
diagonal disorder in graphene are similar to the observed for
conventional quantum Hall systems with diagonal disorder
models,\cite{ando74} but are in clear contrast to the increase of
$\Gamma_n$ with $n$ observed here.

\section{Localization properties}
\label{sec:localization}

While in Sec. \ref{sec:spectrum} we considered the DOS and analyzed
the LL widths as a function of several parameters, we now investigate
the localization properties of states within the LLs. To infer the
degree of localization of the states we use the participation ratio
(PR), which is defined as \cite{thouless74}
\begin{equation}
{\rm PR} = \frac{1}{N' \sum_{i=1}^{N'}|\psi_{i}|^{4} },
\end{equation}
where $\psi_{i}$ is the amplitude of the normalized wave function on
site $i$ and $N'=M\times N$ is the total number of lattice sites. The
PR is therefore directly related to the proportion of the lattice
sites over which the wave function is spread: the PR for a localized
state vanishes in the thermodynamic limit, while peaks in the PR
indicate the presence of extended states (critical energies).

\begin{figure}[t]
\includegraphics[width=\columnwidth]{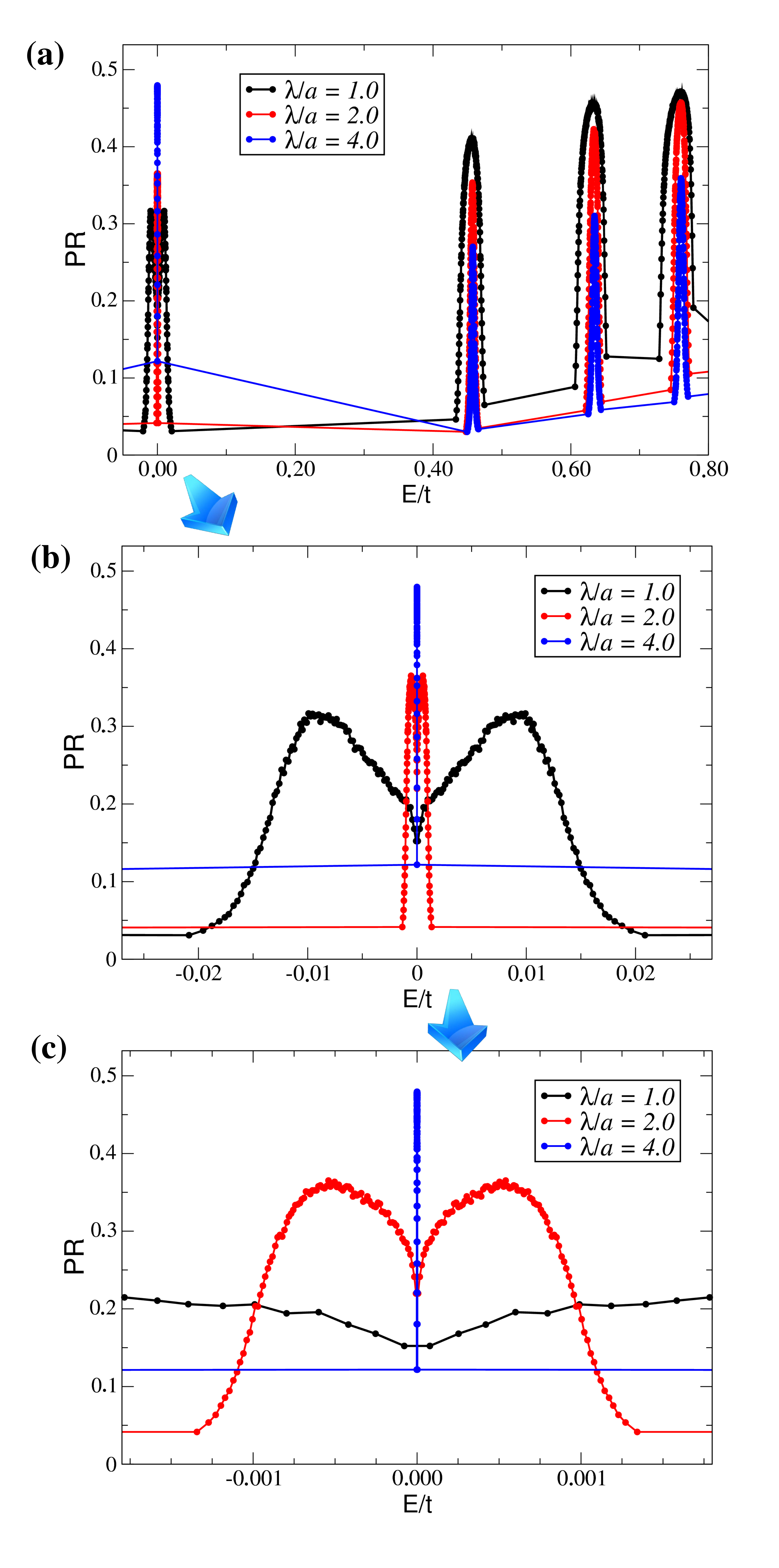}\\
\caption{Participation Ratio as a function of $E/t$ for different
  random hopping correlation lengths $\lambda/a$. Here ,$\Phi /
  \Phi_0=0.02$, $W/t=0.3$, and $M$$\times$$N$=100$\times$90 (namely,
  the same magnetic flux, disorder amplitude, and lattice sizes
  considered in Fig. \ref{fig:LLwidth}). (a) Energy window containing
  the four lowest LLs, $n$=0, 1, 2, and 3. (b) Zoom on the $n=0$
  Landau level. (c) Amplified zoom on the $n=0$ states.}
\label{fig:PR-LLs}
\end{figure}

Figure \ref{fig:PR-LLs}a shows the calculated PR in an energy window
comprising the Landau levels $n$=0, 1, 2, and 3. While the PR for the
$n>0$ LLs indicates the presence of localized states at Landau band
tails and extended states at the band middle, as expected, the PR of
the states from the $n=0$ LL shows a double hump structure, i.e., a
splitting into two peaks. The splitting is more clearly observed in
the energy scale zooms of Fig. \ref{fig:PR-LLs}b and c. This splitting
of critical energies in the lowest LL was previously observed and
discussed for uncorrelated random hopping disorder
\cite{koshino07,pereira09} (and for the similar case of uncorrelated
random magnetic flux disorder \cite{schweitzer08,schweitzer09}). The
novel and interesting aspect we observe here is that the splitting is
rather robust and survives even the sharp width reduction of the $n$=0
LL due to the increasing correlation length. While previous works
considered correlated random hopping disorder,
\cite{kawarabayashi09a,kawarabayashi10} they did not calculate the
localization properties of the states within LLs and therefore missed
this splitting of critical states at $n=0$.

Although experiments \cite{jiang07b} have observed this splitting
energy in the $n$=0 LL, it is hard to make
a quantitative comparison due to the lattice size compromises we need
to make in our simulations. Moreover, we believe that a full
explanation of the experimentally observed splitting requires taking
into account the Zeeman term and possibly electron-electron
interactions, which calls for further theoretical investigation.

\begin{figure}[t]
\includegraphics[width=\columnwidth]{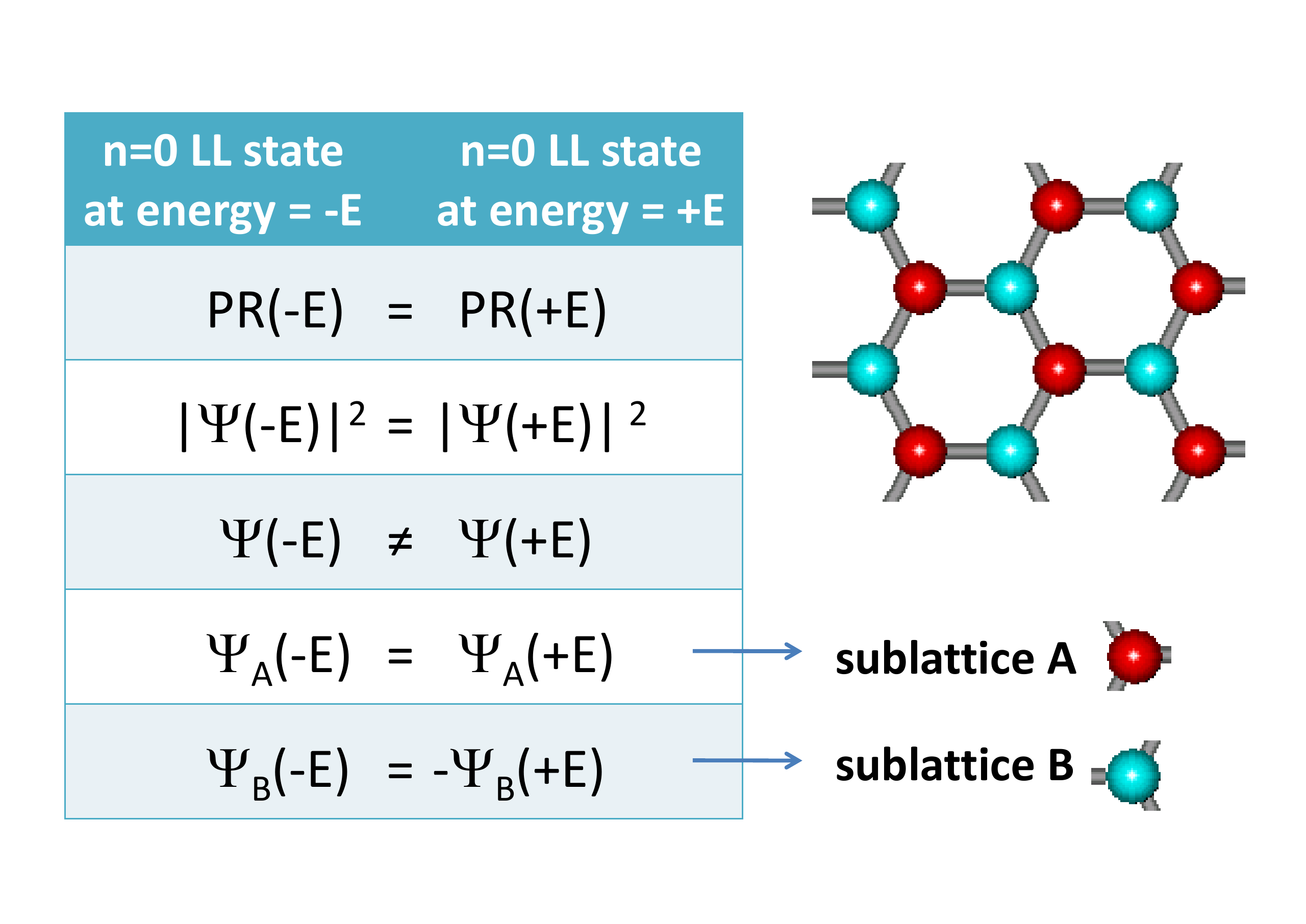}\\
\vskip-0.3cm
\caption{Properties of the wave functions $\Psi$ for $n=0$ LL states
  at energies $+E$ and $-E$. While the PR and the probability
  densities of the wave functions $|\Psi(E)|^{2}$ are identical for
  states at energies $-E$ and $+E$, the wave function amplitudes
  $\Psi(E)$ show differences in only one of the sublattices, in
  analogy to bonding-antibonding states (see text).}
\label{fig7}
\end{figure}

Due to particle-hole symmetry, the probability densities
$|\Psi(E)|^{2}$ and the PR of states at energies $-E$ and $+E$ are
identical (see, for instance, Ref.\ \onlinecite{abergel10} and
references therein). However, when looking directly at the wave function
amplitudes $\Psi(E)$, one can see a difference between the split
states at $-E$ and $+E$: In one of the sublattices (sublattice $A$,
for example) the amplitudes are exactly the same, while in the other
sublattice ($B$) they have the same magnitude but opposite signs, that
is, $\Psi_{A}(-E)=\Psi_{A}(+E)$ and $\Psi_{B}(-E)=-\Psi_{B}(+E)$.
This guarantees that conjugated particle-hole states are orthogonal
and have the same probability densities $|\Psi(E)|^{2}$. It is
noteworthy that orthogonality imposes that the probability weight at
both sublattices is the same. The split states are therefore similar
to bonding-antibonding states. The table in Fig. \ref{fig7} summarizes
these features.

Another feature observed in Fig.~\ref{fig:PR-LLs} is the effect of
increasing correlation length. The effect on the $n$=0 LL is
completely different (opposite) to what is observed in higher
levels. In higher levels, the increase in $\lambda$ causes an overall
reduction of PR values, while in the $n$=0 LL, despite the suppression
of broadening (see Sec. \ref{sec:spectrum}), one observes an overall
increase in the PR. This difference in behavior is probably related to
the fact that the states from the higher levels have much longer
localization lengths than those from the states with $n$=0.
\cite{ando74} To investigate possible finite-size effects on the $n=0$
LL when $\lambda/a$ is large, we ran simulations with larger lattice
sizes (not shown). We found no change in the position of the
identified critical states, as well as no change in the LL widths.

\section{Conclusions}
\label{sec:conclusions}

We have investigated the effects of spatially correlated random
hopping disorder on the structure of Landau levels in graphene. We
quantified the behavior of the $n$th Landau level width $\Gamma_n$ as
a function of the correlation length, as well as other relevant
parameters of the system: disorder amplitude, magnetic field, and
Landau level index $n$. We found that $\Gamma_n$ gets narrower with
increasing correlation length for all Landau levels, and not only for
the $n=0$ level. However, a logarithmic plot of $\Gamma_n$ as function
of $\lambda$ (Fig. 3b) clearly showed that while for $n>0$ the widths
decrease slowly, for $n=0$ they decay exponentially with increasing
$\lambda$. We found no sign of any abrupt vanishing of the $n=0$
Landau level width at finite $\lambda$. This suggests that a different
physical mechanism is behind the narrowing of the $n=0$ LL when
compared to all the other levels (specially when the correlation
length $\lambda$ becomes of the same order of magnitude or higher than
the magnetic length $\ell_B$). We speculate that the underlying
mechanism is due to valley mixing, remnant of the crossover regime.

$\Gamma_n$ increases linearly with both magnetic field and disorder
amplitude. Another interesting observation is that for any fixed
correlation length, $\Gamma_n$ always increases with increasing LL
index $n$, which is completely different to the dependence observed
for diagonal (on-site) disorder models. More importantly, we observe
that the ratio between the n=1 and n=0 Landau level widths,
$\Gamma_1/\Gamma_0$, depends only on the correlation length $\lambda$
and is rather insensitive to the disorder strength and to the
magnitude of the applied magnetic field. This allows a closer contact
of our results with experiments.
In Ref. \ \onlinecite{giesbers09}, for instance, the authors experimentally observe
a $n=0 LL$ width of about 20K, while the width of higher levels is observed to be
about 400K Ð that is $\Gamma_1/\Gamma_0=20$.	Using this information and the results of our
Fig. \ref{fig:universal}, we infer that $\lambda/a=2.2$. This estimate gives a lower bound for
the corrugation length of graphene in the experiment of Ref.\ \onlinecite{giesbers09},
since it neglects all disorder broadening sources but hopping disorder.


We also considered the role played by changing the hopping disorder
correlation length on the localization properties of the states within
the different Landau levels. The splitting of two critical energies
for the $n=0$ level, previously reported for uncorrelated random
hoppings,\cite{pereira09} is still clearly defined for correlated
hopping disorder. It is a robust effect even with the sharp width
reduction of the $n=0$ level that occurs for large values of
$\lambda$. The $n=0$ level can therefore be considered as a
superposition of two levels with broken degeneracy, symmetrically
split around $E=0$. In addition, after analyzing the wave functions of
states belonging to the $n=0$ level, we were able to identify a
symmetric structure of these states. We found that although any two
states at $-E$ and $+E$ have the same probability density and also the
same participation ratio, in one sublattice the wave function
amplitudes of states with energies $\pm E$ are exactly the same, while
in the other sublattice these states have the same magnitude but
opposite signs. Therefore, the origin for this $n=0$ level splitting
is clearly related to the breaking of the sublattice degeneracy
induced by the hopping disorder, a further manifestation of the
influence of valley mixing disorder.


\acknowledgments This work has been supported by the Brazilian funding
agencies FAPERJ, FAPESP, and CNPq. E.R.M. acknowledges partial financial
support from the NSF award DMR 1006230.


\bibliography{graphene2,DiracFermions,QHE}

\end{document}